\documentclass[epj]{webofc}
\usepackage[varg]{txfonts}   
\usepackage{graphicx}
\usepackage{amsmath}
\usepackage{amsfonts}
\usepackage{amssymb}

\woctitle{MESON2014 - the 13$^\textrm{th}$ International Workshop on Meson Production, Properties and Interaction}
\begin{document}
\selectlanguage{english}
\title{Direct vs. final state tensor meson photoproduction - amplitude analysis}

\author{{\L}ukasz Bibrzycki \inst{1}\fnsep\thanks{\email{lukasz.bibrzycki@ifj.edu.pl}} \and 
        Robert Kami\'nski\inst{2}
}

\institute{Academy of Business in D\k{a}browa G\'ornicza, ul. Cieplaka 1C, 41-300 D\k{a}browa G\'ornicza, Poland
\and
           Institute of Nuclear Physics, Polish Academy of Sciences, Division of Theoretical Physics, 31-342
           Krak\'ow, Poland 
          }

\abstract{ Tensor meson photoproduction is described as either a direct production process or a consequence of the 
final state $\pi\pi$ interactions. We calculate the mass distributions for selected partial waves and confront our
 predictions with the measurements of the CLAS experiment. We also point out the structures in the photoproduction 
 amplitudes which may result in observable effects able to indicate the dominant tensor meson photoporduction mechanism.
}
\maketitle
\section{Introduction}
\label{intro}
Spin 2 resonances, in particular the $f_2(1270)$, are important components of the photoproduced $\pi\pi$ and 
$K\overline{K}$ spectra for effective 
masses above 1 GeV. Thus any comprehensible partial wave analysis in this kinematic region requires the knowledge 
of the tensor meson photoproduction amplitudes. They are also important for analyses of the tensor 
glueball photoproduction as the tensor glueball mixes with conventional isocalar tensor mesons.
\section{Structure of the amplitudes}
\label{struct-of-ampl}
\begin{figure}[ht]
    \centering
    \sidecaption
    \includegraphics[scale=.55]{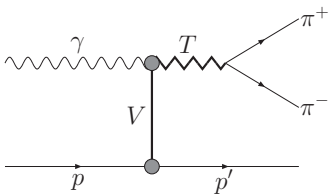}
    \caption{Direct photoproduction diagram}
    \label{direct}     
    \end{figure}

\begin{figure}[h]
    \centering
    \sidecaption
    \includegraphics[scale=.65]{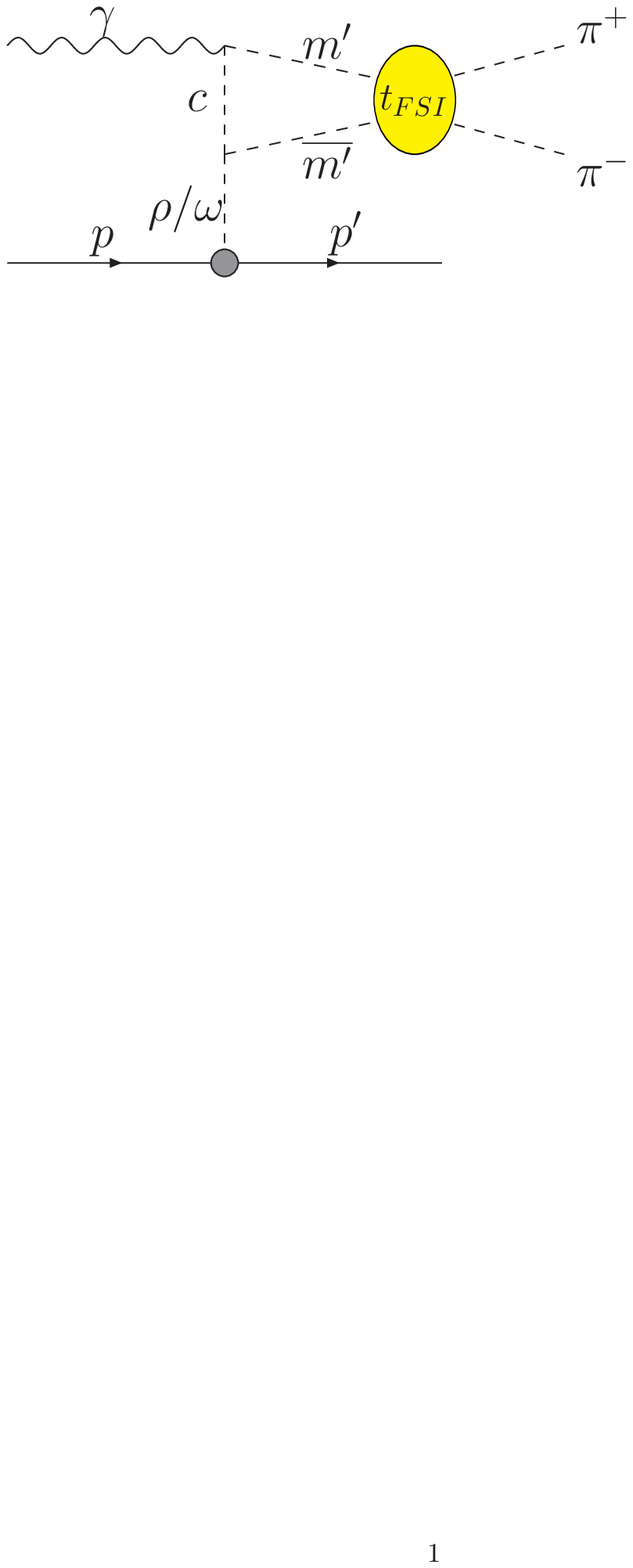}
    \caption{FSI photoproduction diagram}
    \label{FSI}
    \end{figure}

    We consider two models for the amplitudes of the tensor meson photoproduction;
    the direct photoproduction model, where the resonance is produced in the direct channel in the 
    compact spatial region, see Fig.\ref{direct} and the final state interaction (FSI) model, Fig. \ref{FSI}.
    As we will analyze the mechanism of the photoproduction through the correlations of the resonance decay 
    products the 
    complete amplitude is split into two parts, the resonance photoproduction amplitude (Eq.\ref{prod_direct}) and 
    decay amplitude (Eq.\ref{decay}).
    \begin{equation}
    A_p(\lambda_f,\lambda_\gamma,\sigma_1,\sigma_2)=\epsilon^{\mu\nu *}(\lambda_f)M_{\mu\nu;\rho}(\sigma_1,
    \sigma_2)\epsilon^\rho(\lambda_\gamma)
    \label{prod_direct}
    \end{equation}
    where:\\
    $\lambda_f,\lambda_\gamma,\sigma_1,\sigma_2$ - tensor meson, photon, initial and final nucleon helicities 
    respectively,
    $\epsilon^{\mu\nu}(\lambda_f)$-polarization tensor of the tensor meson,
    $\epsilon^\rho(\lambda_\gamma)$- photon polarization vector and
    $M_{\mu\nu;\rho}(\sigma_1,\sigma_2)$- hadron current defined as 
    \begin{equation}
    M_{\mu\nu;\rho}(\sigma_1,\sigma_2)=\bar{u}(p_2,\sigma_2)\Gamma^{\alpha}u(p_1,\sigma_1)G_{\alpha\beta}(t)
    A_{\mu\nu;\beta\rho}N_{V\!NN}(t)N_{\gamma VT}(t)
    \end{equation}
    where:\\
    $N_{V\!NN}(t)$ and $N_{\gamma VT}(t)$-form-factors, $\Gamma^\alpha$-$NNV$ vertex function, $G_{\alpha\beta}$ - 
    propagator of the intermediate vector ($\rho$ and $\omega$) meson and
    $A_{\mu\nu;\beta\rho}$-$\gamma T V$ vertex function which after application of the vector 
    meson and tensor meson dominance \cite{Ren} is determined by 1 coupling constant.
    
    The tensor meson decay amplitude
    \begin{equation}
    A_d(\lambda_f)=\frac{G_{f\pi\pi}}{M_f}\epsilon^{\rho\sigma}(\lambda_f)(k_1-k_2)_\rho(k_1-k_2)_\sigma
    \label{decay}
    \end{equation}
    is defined in terms of: $f_2\pi\pi$ coupling constant - $G_{f\pi\pi}$, $f_2(1270)$ mass - $M_f$ and
    the 4-momenta of the $\pi^+$ and $\pi^-$ - $k_1, k_2$.
    
    In the final state interaction (FSI) mechanism initially 2 pions are produced from the 
    diffuse spatial region. This reaction is described in terms of the $D$-wave projected Born amplitude. Pions 
    then undergo final state interactions which may result in the $D-$wave resonance production \cite{BiKa}, see 
    Fig.\ref{FSI}.
    The general form of the FSI amplitude is:
    \vspace{-.4cm}
    \begin{multline}
     A_{\pi^+\pi^-}(\lambda_f,\lambda_\gamma,\sigma_1,\sigma_2)=
    V_{\pi^+\pi^-}(\lambda_f,\lambda_\gamma,\sigma_1,\sigma_2)+\\
    4\pi\!\!\sum_{m'\overline{m}'}\int_0^\infty\frac{\kappa'^2d\kappa'}{(2\pi)^3}F(\kappa,\kappa')\langle 
    \pi^+\pi^- 
    |\hat{t}_{FSI}|m'\overline{m}'\rangle G_{m'\overline{m}'}(\kappa')
    V_{m'\overline{m}}(\lambda_f,\lambda_\gamma,\sigma_1,\sigma_2)
    \end{multline}
    where:\\
    $V_{m'\overline{m}}$-Born amplitude, $t_{FSI}$-final state scattering amplitude \cite{G-M_K_P_R-E_Y},
    $G_{m'\overline{m}'}(\kappa')$-propagator of the intermediate meson pair,
    $F(\kappa,\kappa')$- form-factor regularizing the meson loop.
    In our calculations the kinematical variables are defined in the s channel helicity frame.

Direct photoproduction amplitude can be expressed in terms of the nucleon spin matrices (Eq.\ref{spin_str})    
\begin{equation}
   A(\sigma'\sigma)\sim a \vec{\epsilon}\cdot
   \vec{\Gamma}_{\sigma'\sigma}\left[(\vec{q}\cdot\vec{k})^2-\frac{1}{3}\vec{k}^2\vec{q}^{\,2}\right]+
   b (q\cdot\Gamma_{\sigma'\sigma}) (\vec{\epsilon}\cdot\vec{k}) (\vec{q}\cdot\vec{k})
   +c\left[(\vec{\epsilon}\cdot\vec{k})(\vec{\Gamma}_{\sigma'\sigma}\cdot\vec{k})-
   \frac{1}{3}\vec{k}^2
   (\vec{\epsilon}\cdot\vec{\Gamma}_{\sigma'\sigma}) \right]
   \label{spin_str}     
\end{equation}
   where: $q, k$ - photon and outgoing pion 4-momenta, $\Gamma_{\sigma'\sigma}$ - spin matrix of the 
   nucleon vertex.
   Important observation concerning the spin structure of this amplitude 
   is that the third term of (Eq.~\ref{spin_str}) is responsible for spin correlation between the tensor meson and the 
proton which can be detected provided the polarization of the proton is measured. There are no terms of 
that kind in the FSI model.

\section{Results}
\label{results}
  We compared the predictions of both models with the $f_2(1270)$ photoproduction data on the mass 
  distribution and helicity projected mass distributions from the CLAS experiment 
  \cite{Bat}.
  \begin{figure}
  \sidecaption
  \centering
  \includegraphics[scale=.3, clip]{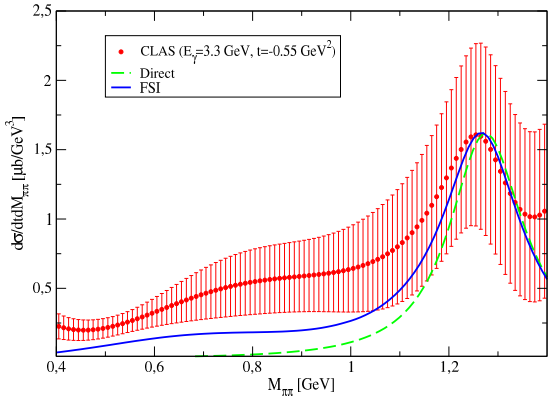}
  \caption{The $D$-wave mass distribution predicted by the direct and FSI photoproduction models compared to CLAS data} 
  \end{figure}
  The models predict similar values of the helicity projected mass distributions but the direct photoproduction 
  model requires the background amplitudes to be added in order to describe the mass distribution.
  \vspace{-.4cm}
  \begin{figure}
  \sidecaption
  \centering
  \includegraphics[scale=.3, clip]{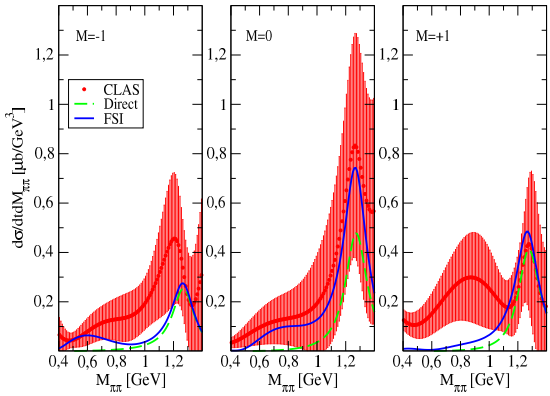}
  \caption{Predictions of both models compared with mass distributions projected on -1, 0 and +1 partial waves measured by CLAS} 
  \end{figure}
\section{Summary}
\label{summary}
Both models agree that the $f_2(1270)$ photoproduction is dominated by the $\lambda_f$=0 partial wave.
The measurement of spin correlations between the final tensor meson and the proton may be used to indicate the dominant mechanism of the tensor meson photoproduction.
\begin{acknowledgement}
This work has been partly supported by the National Science Centre (grant No UMO-2013/09/B/ST2/04382).
\end{acknowledgement}


\begin{thebibliography}{5}
\bibitem{Ren} B. Renner, Nucl. Phys. B \textbf{30}, 634 (1971)
\bibitem{BiKa} \L. Bibrzycki, R. Kami\'nski, Phys. Rev. D \textbf{87}, 114010 (2013)
\bibitem{G-M_K_P_R-E_Y} R. Garcia-Martin \textit{et al.} Phys. Rev. D \textbf{83}, 074004 (2011)
\bibitem{Bat} M. Battaglieri \textit{et al.}, Phys. Rev. D \textbf{80}, 072005 (2009)
\end{thebibliography}
\end{document}